      [1999/12/01 v1.4c Il Nuovo Cimento]
\documentclass[varenna]{cimento}

\usepackage{graphicx}  
\title{Deformation and the Nuclear Matrix Elements of the 
Neutrinoless $\beta\beta$ Decay}
\author{J. Men\'endez \atque \underline{A. Poves}}
\institute{Departamento de F\'isica Te\'orica and IFT-UAM/CSIC, Universidad~Aut\'onoma~de~Madrid,
E-28049, Madrid, Spain}

\author{E. Caurier \atque F. Nowacki}
\institute{IPHC, IN2P3-CNRS/Universit\'e Louis Pasteur BP 28, F-67037, Strasbourg~Cedex~2, France}

\shortauthor{Men\'endez, Poves, Caurier \atque Nowacki}

\begin{document}

\maketitle

\begin{abstract}In this talk I will review the ``state of the art'' of the calculations of
the nuclear matrix elements (NME) of the
neutrinoless double beta decays ($0\nu\beta\beta$) for the nuclei
$^{48}$Ca, $^{76}$Ge, $^{82}$Se, $^{124}$Sn, $^{128}$Te, $^{130}$Te and $^{136}$Xe
in the framework of the Interacting Shell Model (ISM), and compare them with the
NME's obtained  using the Quasi-particle RPA approach (QRPA). I will also discuss the
effect of the competition between the pairing and quadrupole correlations in the
value of these NME's. In particular I will show that, as the difference in deformation
between parent and grand daughter grows, the NME's of both the neutrinoless and the
two neutrino modes decrease rapidly.
\end{abstract}

\section{Introduction}

The discovery of neutrino oscillations in recent experiments at
Super-Kamiokande \cite{PhysRevLett.81.1562}, SNO
\cite{PhysRevLett.89.011301} and KamLAND \cite{PhysRevLett.90.021802}
has changed the old conception of neutrinos by proving that they are
massive particles. According to the origin of their mass, neutrinos
can be either Dirac or Majorana particles, the latter case being
particularly interesting since it would imply an extension to the
standard model of electroweak interactions, and, being neutrinos their
own antiparticles in this scenario, lepton number conservation would be
broken. Besides, it happens that the best way to detect one of these
violating processes and consequently establish the Majorana character
of the neutrinos would be detection of the neutrinoless double beta decay
($0\nu\beta\beta$).

Double beta decay is a very slow weak process. It takes place between
two even-even isobars when the single beta decay is energetically
forbidden or hindered by large spin difference. Two neutrinos beta
decay is a second order weak process ---the reason of its low rate---,
and has been measured in a few nuclei. The $0\nu\beta\beta$ decay
is analog but needs neutrinos to be Majorana particles. With the exception
of one unconfirmed claim \cite{KlapdorKleingrothaus:2001ke,KlapdorKleingrothaus:2004wj},
it has never been observed, and currently there is a number of experiments
either taking place \cite{arnold:182302,arnaboldi:142501,bloxham:025501}
or expected for the near future ---see e.g. ref. \cite{Avignone:2007fu}---
devoted to detect this processes and to set up firmly the nature of
neutrinos.

Furthermore, $0\nu\beta\beta$ decay is also sensitive to the absolute
scale of neutrino mass, and hence to the mass hierarchy ---at present,
only the difference between different mass eigenstates is known. Since
the half-life of the decay is determined, together with the masses,
by the nuclear matrix element (NME) for this process, the knowledge
of these NME's is essential to predict the most favorable decays and,
once detection is achieved, to settle the neutrino mass scale and
hierarchy.

Two different and complementary methods are mainly used to calculate
NME's for $0\nu\beta\beta$ decays. One is the family of the quasiparticle
random-phase approximation (QRPA). This method has been used by different
groups and a variety of techniques is employed, with results for most
of the possible emitters \cite{Suhonen:1998ck,Rodin:2006yk,Rodin:2007fz}. This
work concerns to the alternative, the interacting shell model (ISM) \cite{Caurier:2004gf}. 

In previous works \cite{Retamosa:1995xt,Caurier:prl.77.1954}, the
NME's for the $0\nu\beta\beta$ decay were calculated taking into account
only the dominant terms of the nucleon current. However, in ref. \cite{Simkovic:1999re}
it was noted that the higher order contributions to the current (HOC) are
not negligible and it was claimed that they could reduce up to 20\%-30\%
the final NME's. Subsequently, other QRPA calculations \cite{Kortelainen:2007rh,Kortelainen:2007mn}
have also taken into account these terms, although resulting in a
somewhat smaller correction. These additional nucleon current contributions
have been recently included for the first time in the ISM framework
\cite{Caurier:2007wq}.
In addition, the short range correlations (SRC),  are now modeled
either by the Jastrow prescription or by the UCOM method \cite{Feldmeier:1997zh}.

\section{ISM $vs$ QRPA Nuclear Matrix Elements}
The expression for the half-life of the
$0\nu\beta\beta$ decay can be written as \cite{Doi:1980ze,Doi:1985dx}:

\begin{equation}
\left(T_{1/2}^{0\nu\beta\beta}\left(0^{+}\rightarrow0^{+}\right)\right)^{-1}=G_{01}\left|M^{0\nu\beta\beta}\right|^{2}\left(\frac{\left\langle m_{\nu}\right\rangle }{m_{e}}\right)^{2},\label{eq:t-1}\end{equation}

where $\left\langle m_{\nu}\right\rangle =\sum_{k}U_{ek}^{2}m_{k}$
is the averaged neutrino mass, a combination of the neutrino masses $m_{k}$ 
due to the neutrino mixing matrix $U$ ---as we see, the neutrino
mass scale is directly related to the decay rate--- and $G_{01}$
is a kinematic factor ---dependent on the charge, mass and available
energy of the process. $M^{0\nu\beta\beta}$ is the NME object of
study in this work. 

The kinematic factor $G_{01}$ depends on the value of the coupling constant $g_{A}$. Therefore
we have to take this into account when comparing the values of NME's obtained with different $g_{A}$ values.
In these cases we will use a NME modified as:

\begin{equation}
M'^{\;0\nu\beta\beta}=\left(\frac{g_{A}}{1.25}\right)^{2}M^{0\nu\beta\beta}
\end{equation}

These $M'^{\;0\nu\beta\beta}$'s are directly comparable between them no matter
which was the value of $g_{A}$ employed in their calculation, since they share a common
$G_{01}$ factor ---that of $g_{A}=1.25$. In this sense, the translation of
$M'^{\;0\nu\beta\beta}$'s into half-lives is transparent. The QRPA results obtained with different
$g_{A}$ values are already expressed in this way by the authors of refs. \cite{Simkovic:2007vu,Rodin:2007fz}
while the results of refs. \cite{Kortelainen:2007rh,Kortelainen:2007mn,Suhonen:2008ijmpa} will be translated
by us into the above form when compared with other results.

We have calculated the ISM NME's both taking the  UCOM  and Jastrow
ansatzs  for the short range correlations. The former use 
the correlator of the $ST=01$ channel \cite{Roth:2005pd}, 
throughout the calculation. The correlator of the
other important ---even--- channel is very similar to this one, and it should not make
much difference on  this result.

\begin{figure}

\begin{center}\includegraphics[%
  scale=0.55,
  angle=270]{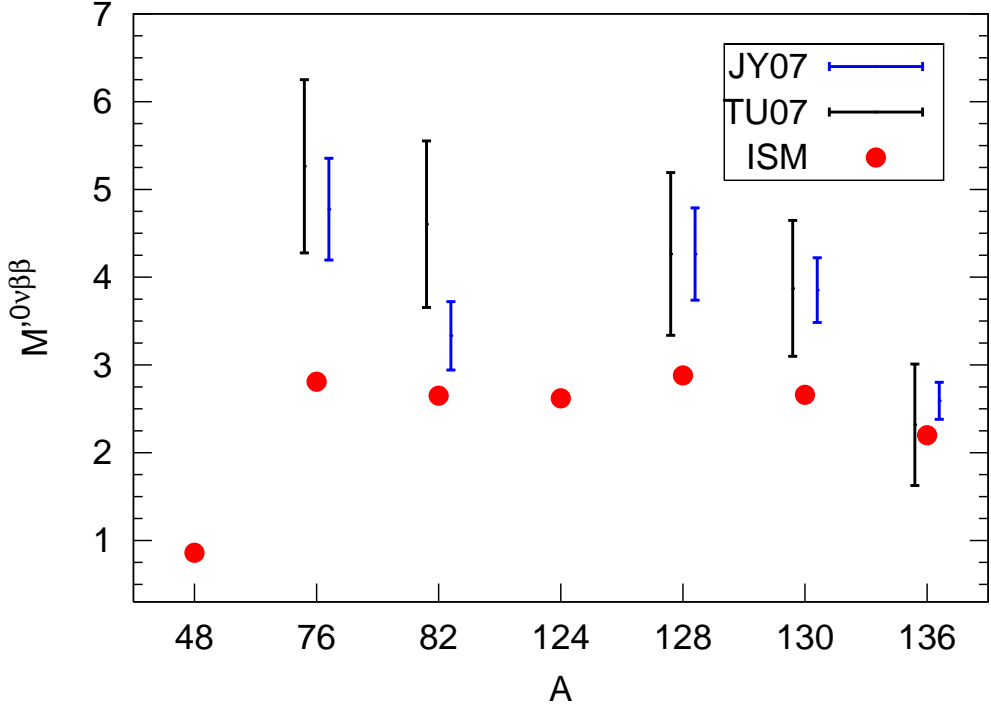}\end{center}

\caption{\label{cap:m0nbb_ucom} The neutrinoless double beta decay $M'^{\;0\nu\beta\beta}$'s for ISM and QRPA 
calculations treating the SRC with the UCOM approach. Tu07 QRPA results from
ref. \protect\cite{Simkovic:2007vu} and Jy07 results from refs. \protect\cite{Kortelainen:2007rh,Kortelainen:2007mn}.}

\end{figure}

\begin{figure}

\begin{center}\includegraphics[%
  scale=0.55,
  angle=270]{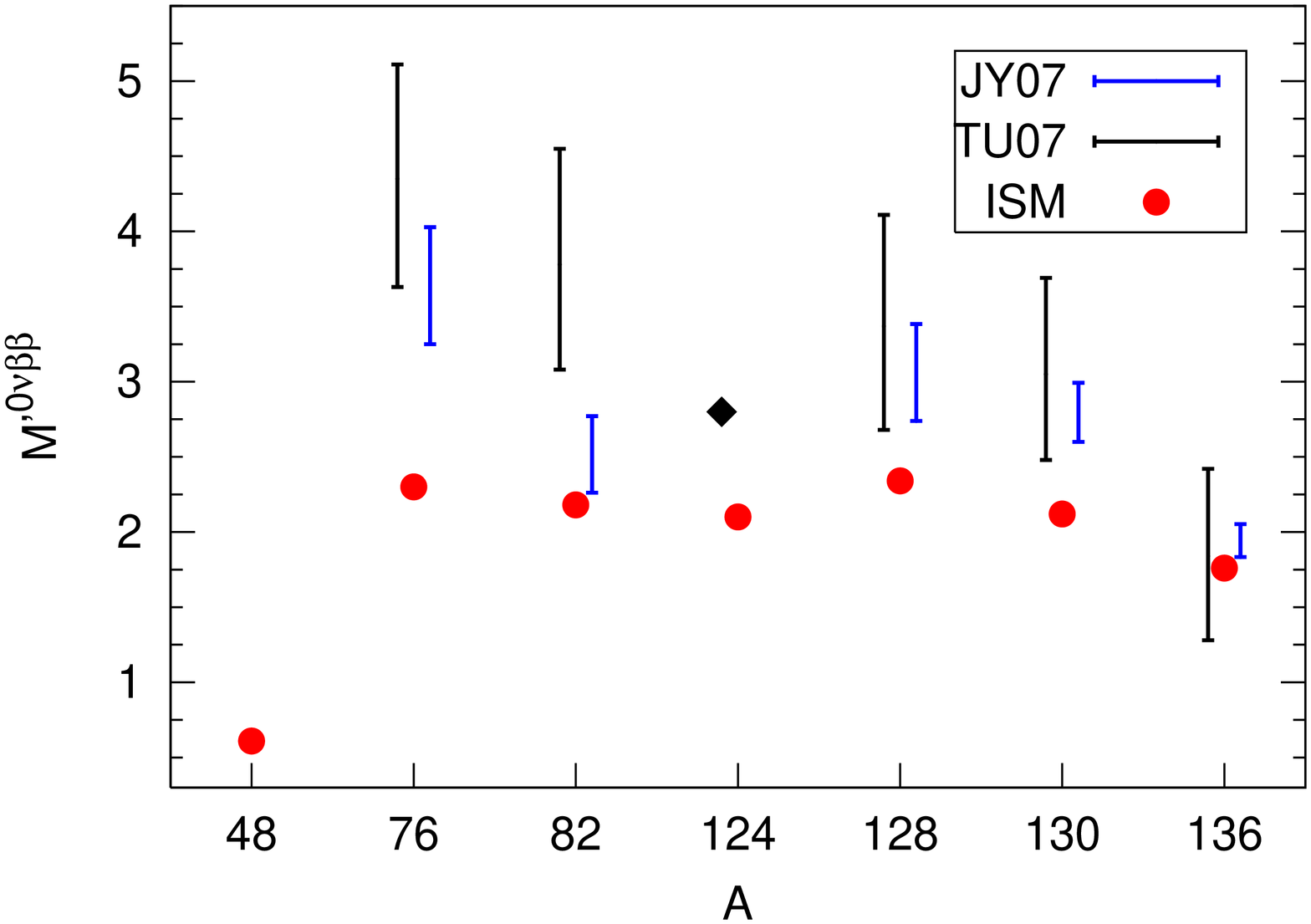}\end{center}

\caption{\label{cap:m0nbb} Same as fig.\ref{cap:m0nbb_ucom} but with Jastrow type SRC. Tu07 QRPA results from 
ref. \protect\cite{Rodin:2007fz} and Jy08 results from ref. \protect\cite{Suhonen:2008ijmpa}.}

\end{figure}

In figure \ref{cap:m0nbb_ucom} the ISM and QRPA results for the NME's 
are compared within this UCOM treatment of the SRC. The same figure but considering
Jastrow type SRC was shown in ref. \cite{Caurier:2007wq}, but, inadvertently, the QRPA 
results from refs. \cite{Kortelainen:2007rh,Kortelainen:2007mn,Suhonen:2008ijmpa}  obtained
with $g_A$=1.0  where not transformed properly.
This is corrected in figure \ref{cap:m0nbb}.
By comparing both figures, it is confirmed that
there is a common trend; when the nuclei that participate in the decay have  a low
level of quadrupole correlations, as in the decays of $^{124}$Sn and  $^{136}$Xe,
 both approaches agree.  The QRPA in a spherical basis seems not to be able to
capture the totality of the quadrupole correlations when they are strong. As these correlations
tend to reduce the NME's, the QRPA produces NME's that are too large in
$^{76}$Ge, $^{82}$Se,  $^{128}$Te, and $^{130}$Te. For both ISM and QRPA 
the only net effect of UCOM is an increase of the Jastrow results of about 20\%.


\section{The Influence of Deformation in the NME's }

An important issue regarding $0\nu\beta\beta$ decay is the role
of pairing and deformation. It has recently been discussed in ref.
\cite{Caurier:2007wq} that the pairing interaction favors the $0\nu\beta\beta$
decay and that, consequently, truncations in seniority, not including
the anti-pairing-like effect of the missing uncoupled pairs, tend
to overestimate the value of the NME's. On the other hand, the NME is also
reduced when the parent and grand-daughter nuclei have different deformations
\cite{Caurier:2007qn}. Thus, we have studied the interplay
between both pairing and deformation, this is, to which extent a wave
function in the laboratory frame, truncated in seniority, can capture
the correlations induced by the quadrupole-quadrupole part of the
nuclear interaction, and its eventual influence in $0\nu\beta\beta$
NME's. There is an extra motivation to pursue this study; the possibility
to carry on an experiment with $^{150}$Nd, which is a well deformed nuclei,
decaying into $^{150}$Sm which is a much less deformed one.

To study the interplay between pairing, seniority truncations,
and quadrupole correlations we need first to decide how to measure
the quadrupole correlations of the ground state. Our choice is to
refer to non energy-weighted sum rule:

\begin{eqnarray}
\left\langle Q^{2}\right\rangle  & = & \sum_{i}\left|\left\langle 2_{i}^{+}\right|Q\left|0^{+}\right\rangle \right|^{2}\label{eq:q2_def}\end{eqnarray}

\begin{figure}

\begin{center}\includegraphics[%
  scale=0.55]{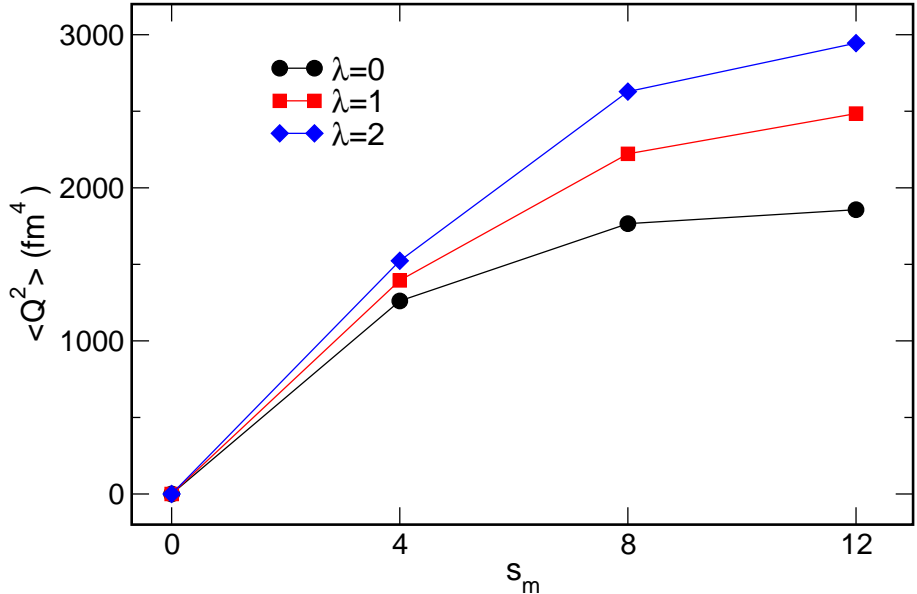}\end{center}

\caption{\label{cap:Q_lambda_senior}Quadrupole correlations in the ground
state of $^{82}$Kr as a function of the amount of quadrupole-quadrupole
interaction $\lambda Q\cdot Q$ added to the hamiltonian and of the maximum seniority
$s_m$ permitted in the wave functions}.

\end{figure}

The operator $Q$ represents the mass quadrupole. No effective ``nuclear'' charges
are included.
Using $^{82}$Kr as our test bench, we proceed to compute $\left\langle Q^{2}\right\rangle $,
first with our standing effective interaction and different seniority
truncations (s$_m$ means the maximum seniority allowed in the wave 
functions of parent and grand daughter nuclei).
  The results are drawn in Figure \ref{cap:Q_lambda_senior}
as the black circles labeled $\lambda=0$. We can see that at $s_m=4$
---roughly, the implicit level of seniority truncation in the spherical QRPA--- 
some 70\% of the full quadrupole correlations
are incorporated in the wave function. We would like to know how this
behavior evolves when more correlations are enforced in the system.
For this we recalculate the ground state of $^{82}$Kr with a new
hamiltonian that consists of the standing one plus a quadrupole-quadrupole
term $\lambda Q\cdot Q$, whose effect will be gauged by its influence
in the sum rule. To have an idea of the relevant range of values of
$\left\langle Q^{2}\right\rangle $ in this nucleus and valence space,
we have gone to the limit of pure quadrupole-quadrupole interaction
with degenerate single particle energies, getting $\left\langle Q^{2}\right\rangle \approx 4500$
fm$^{4}$. The results for $\lambda=1$ and $\lambda=2$
are also shown in Figure \ref{cap:Q_lambda_senior} ($\lambda=1$ corresponds to
$\lambda_{qq}=0.025$ in Figs. 6-9  and $\lambda_{qq}=1$ in Fig. 10
to $\lambda_{qq}=0.1$. It is evident
in the figure that, as we try to increase the correlations, the $s_m=4$
truncation becomes more and more ineffective. For $\lambda=1$, only
57\% of the full correlations are present, and for $\lambda=2$ only
50\%. The situation is different for $^{82}$Se: while for $\lambda=0$
the values of $\left\langle Q^{2}\right\rangle $ as a function of
seniority are similar, albeit a bit smaller than the $^{82}$Kr ones,
for $\lambda=1$ and $\lambda=2$ there is scarcely any increase of
the ground state correlations. This means also that, as we increase
$\lambda$, the ``deformation'' of $^{82}$Kr grows, whereas that
of $^{82}$Se remains constant.

\begin{figure}

\begin{center}\includegraphics[%
  scale=0.55]{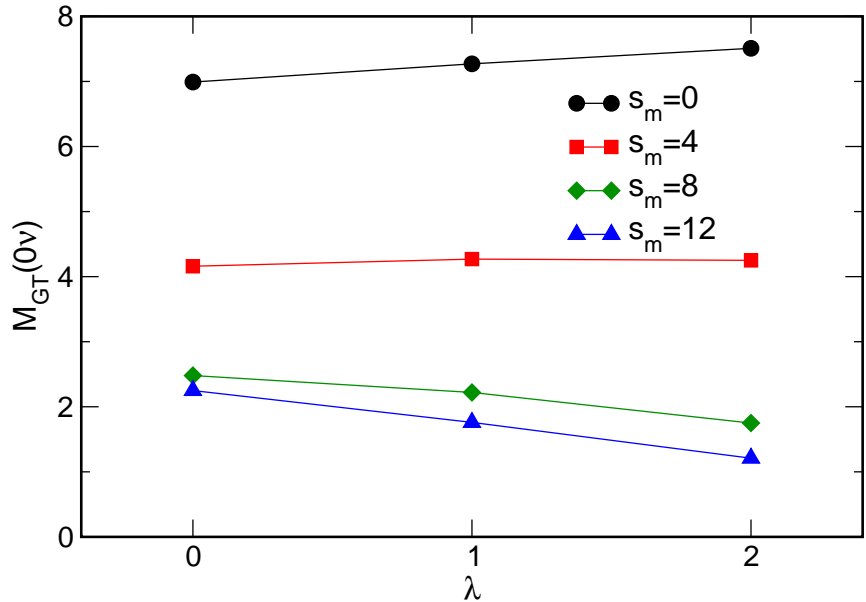}\end{center}

\caption{\label{cap:Mgt_lambda_senior}$^{82}$Se $\rightarrow$ $^{82}$Kr
Gamow-Teller matrix element, $M^{GT}$, as a function of the maximum
seniority of the wave functions, for different values of the strength
of the extra quadrupole-quadrupole interaction.}

\end{figure}

This behavior offers us the opportunity of exploring the effect of
the difference in deformation between parent  and grand daughter in the $0\nu\beta\beta$
NME's. To this goal, we have computed the Gamow-Teller matrix element for
different values of $\lambda$ ---the amount of extra quadrupole-quadrupole
interaction--- and $s_m$ ---the maximum seniority allowed in the
wave function---. The results are gathered in Figure \ref{cap:Mgt_lambda_senior}.
For $s_m=0$, we observe that the Gamow-Teller matrix element grows
as a function of $\lambda$. This may seem paradoxical, but is not,
because at this seniority truncation, the only effect of adding more 
quadrupole-quadrupole
interaction is to augment the pairing content of the wave functions,
thus increasing $M^{GT}$. At $s_m=4$, $M^{GT}$ remains constant
as a function of $\lambda,$ meaning that the minor increase of the
correlations of $^{82}$Kr, that we have shown in Figure \ref{cap:Q_lambda_senior},
is barely enough to compensate the increase of $M^{GT}$ at $s_m=0$.
On the contrary, the full space results are sensitive to the difference
in deformation ---or, to be more precise, to the difference in the
level of quadrupole correlations in the ground state--- between parent
and grand daughter. The effect goes in the direction of reducing the
value of $M^{GT}$. In the $A=82$ decay, doubling the quadrupole
correlations in $^{82}$Kr, roughly halves $M^{GT}$.

\begin{figure}

\begin{center}\includegraphics[%
  scale=0.55]{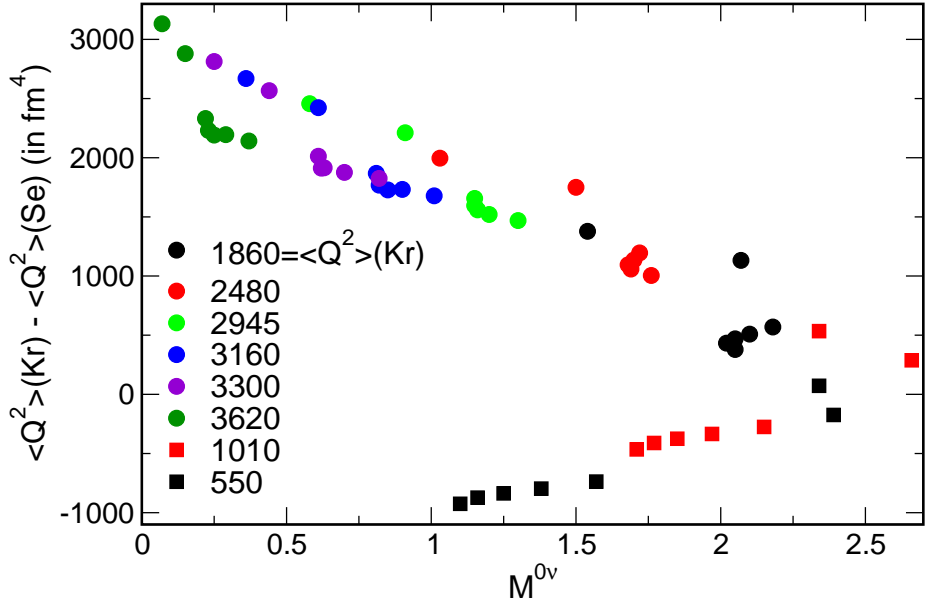}\end{center}

\caption{\label{cap:q20nuX}$^{82}$Se $\rightarrow$ $^{82}$Kr
NME, $M^{0\nu}$, as a function of the difference between the mass
quadrupole sum rule between $^{82}$Kr and $^{82}$Se for a large
number of different values of the strength of the added quadrupole
quadrupole interaction.}

\end{figure}

  We can go much further in the exploration of the deformation effects,
calculating the NME of the decay for initial and final states computed
with different amounts of supplementary quadrupole-quadrupole interaction.
As before , we measure the quadrupole correlations by means of the mass quadrupole sum rule.
The results of this search are plotted in Fig. \ref{cap:q20nuX}. We observe
that the NME decreases almost linearly as the difference of the 
sum rules for the final and initial states increases. In fact, the maximum values
of the NME are reached when this difference is close to zero. On the contrary
for large differences the NME can be extremely quenched. Notice that the 
values in the figure range between 0.07 and 2.7, a factor of 40 span.   
The $\lambda$=0 value is 2.18, indicating that the difference in 
quadrupole correlations between  $^{82}$Kr and $^{82}$Se is not very large.

\section{0$\nu$ (Unphysical) Mirror Decays: A Case Study}

We have also studied the transitions between mirror nuclei in order to have a clearer view of the
role of deformations in the NME's. These transitions have the peculiarity that the wave functions
 of the initial
and final nuclei are identical (provided Coulomb effects are neglected)
and consequently the interplay of the $0\nu\beta\beta$ operator and of the nuclear 
wave functions in the NME may be easier to understand.

We have studied four parent nuclei with six valence protons and four valence neutrons,
$^{26}$Mg, $^{50}$Cr, $^{66}$Ge and $^{110}$Xe, decaying into the four grand daughter nuclei 
with four valence protons and six valence neutrons,
$^{26}$Si, $^{50}$Fe, $^{66}$Se and $^{110}$Ba. The valence spaces considered are the $sd$-shell, the $pf$-shell,
r$_3$g  (1p$_{3/2}$, 1p$_{1/2}$, 0f$_{5/2}$, 0g$_{9/2}$)
and r$_4$h (0g$_{7/2}$, 1d$_{5/2}$, 1d$_{3/2}$, 2s$_{1/2}$, 0h$_{11/2}$). The starting interactions are
USD \cite{Wildenthal:84}, KB3 \cite{Poves:81}, GCN28.50 and GCN50.82 \cite{Gniady:0000aa}.
The deformation of the nuclei is modified 
as in the previous section, and it is quantified also in the same way.

\begin{figure}

\begin{center}\includegraphics[%
  scale=0.55,
  angle=270]{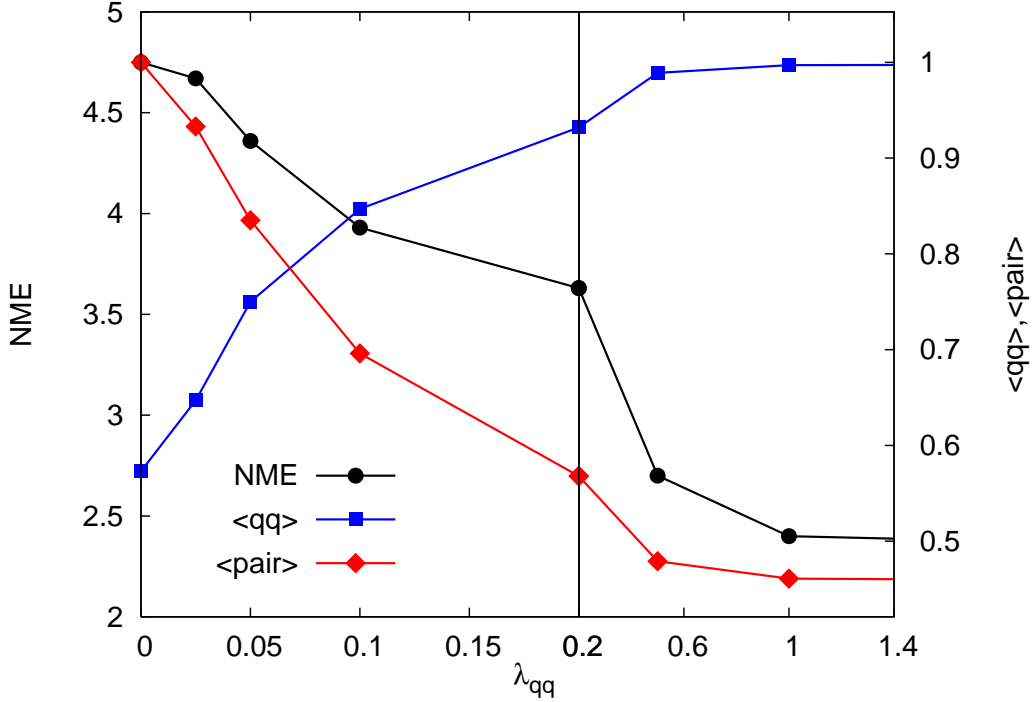}\end{center}

\caption{\label{cap:qq_diag66} $^{66}$Ge $\rightarrow$ $^{66}$Se
NME, $M^{0\nu}$, as a function of the strength of the added quadrupole
quadrupole interaction. Equally deformed case; the same amount of extra QQ  
interaction is added to  $^{66}$Ge and $^{66}$Se. On the right hand y axis
the pairing and quadrupole sum rules are represented, normalized so that 
their maximum value is 1. Note the change of
scale in the x axis at $\lambda=0.2$.}

\end{figure}

\begin{figure}

\begin{center}\includegraphics[%
  scale=0.55,
  angle=270]{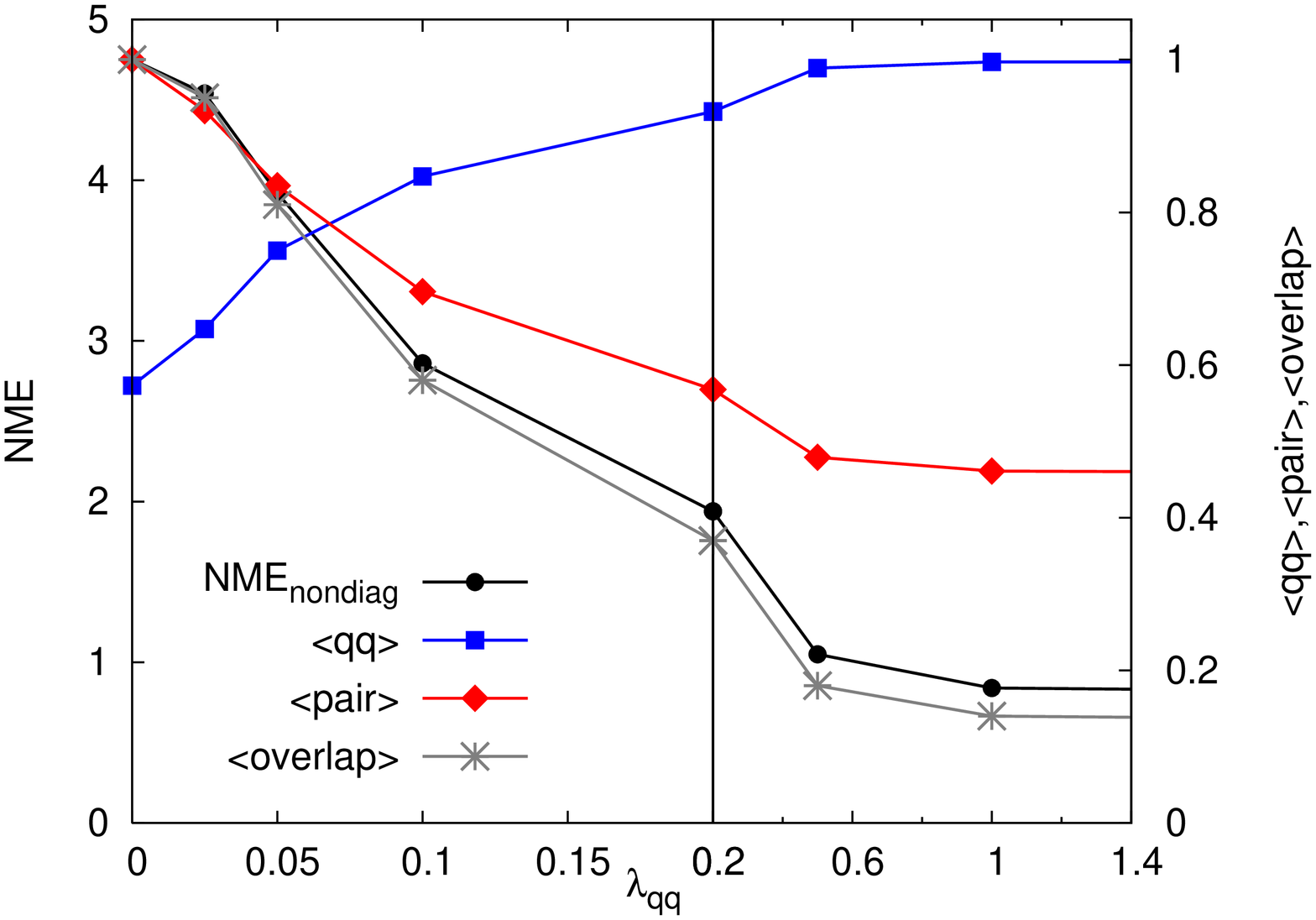}\end{center}

\caption{\label{cap:qq_nondiag66} Same as previous figure, but now the 
additional quadrupole interaction is only added to  $^{66}$Se.
The normalized overlap between the initial and final states is also included.}

\end{figure}

The results for $A=66$ in the case of equally deformed  initial and final nuclei, are shown on 
Fig. \ref{cap:qq_diag66}. There we see that, as the nuclei become more deformed, the
NME and the pairing content of the wave function get smaller, while the quadrupole sum rule grows. All these changes
are nearly linear for reasonable deformations and then the saturation is approached more smoothly. Note that the
purely quadrupole interaction ($\lambda_{qq}$$\rightarrow$$\infty$ limit) 
gives a NME which is about a half of the value obtaind with no additional quadrupole.

Fig. \ref{cap:qq_nondiag66} shows the same quantities than  Fig. \ref{cap:qq_diag66} but now only the final nucleus has been
artificially deformed by adding an extra quadrupole-quadupole term. In addition, the overlap between initial and
final wave functions has been included. We see that now, the reduction of the NME is more pronounced and, what is more
interesting, that it follows closely the overlap between wave functions. This means that, 
if we write the final wave function as: $\left|\Psi_{\,}\right\rangle =a\left|\Psi_{0}\right\rangle +b\left|\Psi_{qq}\right\rangle $,
the $0\nu\beta\beta$ operator only connects the parts of the wave functions that have the same deformation among
themselves.

The behavior of the NME's  with respect to the difference of deformation between parent and grand daughter  
is common to all the other transitions between mirror nuclei that we have studied. Therefore we can submit that
this is a robust result.
  However, when we consider the transitions between equally deformed nuclei, the evolution of the NME's with
the deformation that we have found in A=66 is only shared by the A=110 case. When
the  valence space is a full major oscillator shell ---$A=26$ and $A=50$---
the situation is quite different. Indeed, what is observed is that the NME does not decrease for moderate values of $\lambda$
but remains rather constant until a point ---with large deformation--- where its value increases significantly ---up to 50\%---.
This is due to the fact that, at this point, the major contribution to the NME ceases to come only form the decay of pairs
coupled to $J=0$, since other values like $J=2,4,6$, which usually have a contribution to the NME contrary to that of $J=0$,
reverse sign and grow until being comparable with this contribution, thus resulting in this notorious rise of the NME.
In the $sd$-shell and $pf$-shell cases, the $\lambda_{qq}$$\rightarrow$$\infty$ limit is equivalent to
Elliott's SU(3) limit, and, the fact that both the initial and final nuclei should belong to the same irrep of SU(3) 
may be the reason of the increase of the NME, but, for the moment we have not found a formal explanation.

\section{2$\nu$ (Unphysical) Mirror Decays: A Case Study}

\begin{figure}

\begin{center}\includegraphics[%
  scale=0.55,
  angle=270]{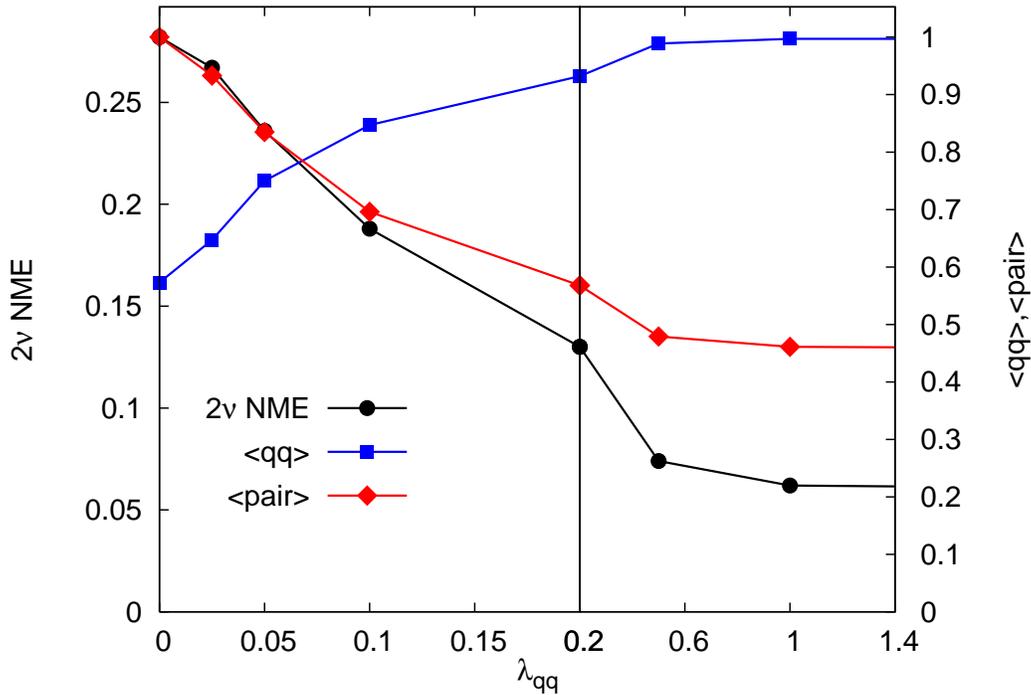}\end{center}

\caption{\label{cap:2nuqq_diag66}Equally deformed $^{66}$Ge $\rightarrow$ $^{66}$Se
$2\nu$NME. Note the change of
scale in the x axis at $\lambda=0.2$.}

\end{figure}

\begin{figure}

\begin{center}\includegraphics[%
  scale=0.55,
  angle=270]{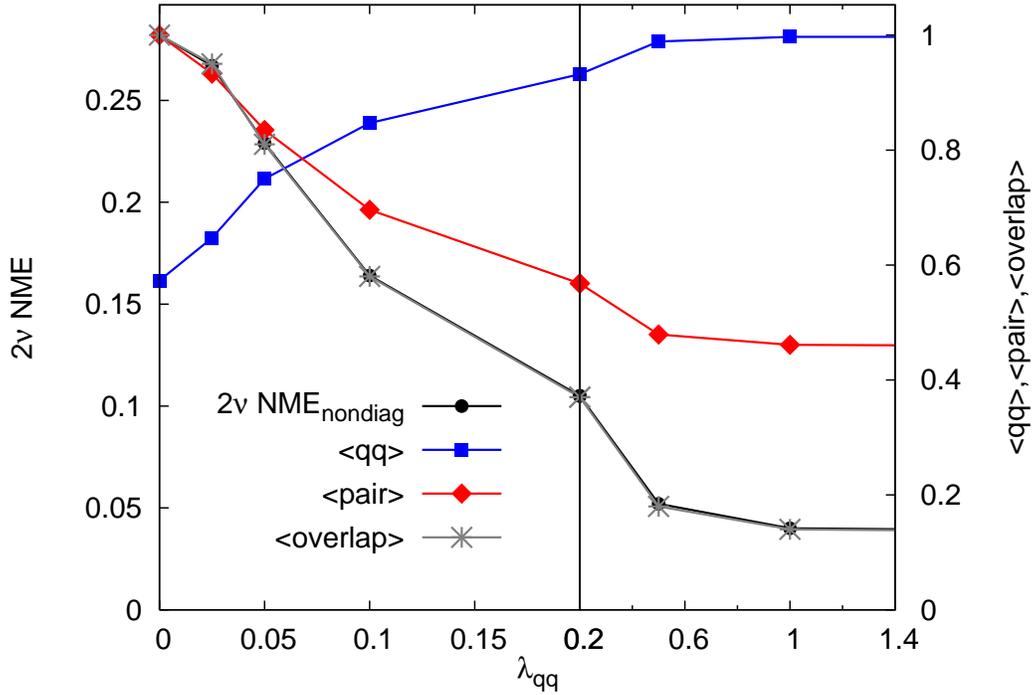}\end{center}

\caption{\label{cap:2nuqq_nondiag66}The same as the previous figure, but now the only nuclei calculated with
additional quadrupole interaction is the final one. The normalized overlap between initial and final
states is also included.}

\end{figure}

Very similar conclusions may be reached for the effect of deformation in $2\nu\beta\beta$ decay. For instance, the
mirror nuclei diagonal and non-diagonal NME's are represented as in the $0\nu\beta\beta$ case in Figs.
 \ref{cap:2nuqq_diag66} and \ref{cap:2nuqq_nondiag66} for the $A=66$ transition. We see that the figures resemble
very much that of the previous section, with the only exception that, in the equally deformed, case, the lowering
of the NME due to deformation is more pronounced.

\begin{figure}

\begin{center}\includegraphics[%
  scale=0.55]{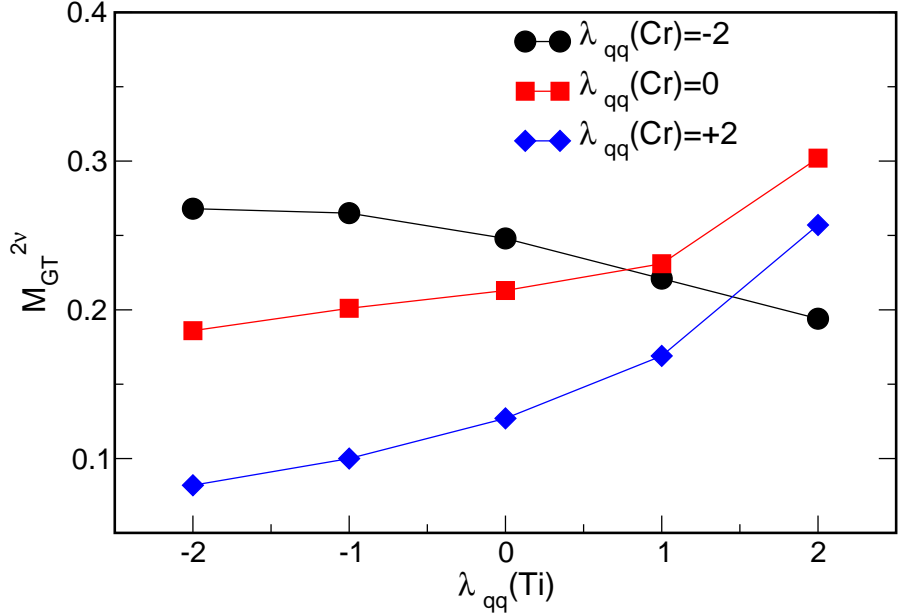}\end{center}

\caption{\label{cap:2nu_ticr} Influence of deformation in the $^{48}$Ti $\rightarrow$ $^{48}$Cr decay.}

\end{figure}

If we look to a non mirror ---but again fictitious--- transition, for instance that of $^{48}$Ti $\rightarrow$ $^{48}$Cr, we get the
results shown in Fig. \ref{cap:2nu_ticr}. If we compare with those of ref. \cite{Caurier:2007qn}, which
are again the equivalent ones for the $0\nu\beta\beta$ transition, we see that there are not substantial changes.
In this sense, deformation seems to affect similarly to $0\nu\beta\beta$ and $2\nu\beta\beta$ decays.

\section{Summary}
After a brief discussion of the ``state of the art'' results for the 
nuclear matrix elements of the neutrinoless double beta decay in the 
context of the Interacting Shell Model and of the Quasiparticle Random  Phase
Approximation, we have analyzed  the role of the pairing correlations
and the deformation in the NME's, concluding that seniority truncations
are less reliable
when the quadrupole correlations are large. Since the NME's  are reduced when
the deformation of parent and grand daughter is different, a bad treatment of the
quadrupole correlations can 
lead to an artificial enhancement of the NME's in the transitions among nuclei
with unequal deformations, or in the cases of nuclei with different --and large-
amounts of quadrupole correlations.

\acknowledgments
This work has been supported by a grant of the Spanish Ministry of
Education and Science, FPA2007-66069,  by the IN2P3-CICyT collaboration agreements,
by the Spanish  Consolider-Ingenio 2010 Program CPAN (CSD2007-00042), and  
by the Comunidad de Madrid (Spain), project HEPHACOS P-ESP-00346.

\bibliographystyle{varenna.bst}

\bibliography{d_beta.bib}

\end{document}